\documentclass[aps,prx,twocolumn,superscriptaddress,showpacs,final,floatfix,longbibliography]{revtex4-2}
\usepackage[utf8]{inputenc}
\usepackage[toc,page]{appendix}
\usepackage{amsfonts}
\usepackage[dvips]{graphicx}
\usepackage{amsmath}
\usepackage{amssymb}
\usepackage{hyperref}
\usepackage{color}
\usepackage{mathrsfs}
\usepackage{isomath}
\usepackage{amsthm}
\usepackage{epstopdf}
\usepackage{txfonts}
\usepackage{dsfont}
\usepackage{ulem}
\usepackage{physics}
\allowdisplaybreaks[4]

\usepackage{bbold}
\usepackage{mathtools}

\newcommand{\ignore}[1]{} 

\newcommand{\hil}{\mathcal{H}}
\newcommand{\obsset}[1]{\boldsymbol{#1}}

\newcommand{\kb}[2]{|#1\rangle\langle #2|} 
\renewcommand{\tr}{\mathrm{tr}} 

\newcommand*{\circled}[1]{\lower.7ex\hbox{\tikz\draw (0pt, 0pt)%
		circle (.5em) node {\makebox[1em][c]{\small #1}};}}

\newcommand{\be}{\begin{equation}}
\newcommand{\ee}{\end{equation}}
\newcommand{\eea}{\end{eqnarray}}
\newcommand{\bea}{\begin{eqnarray}}

\newcommand{\av}[1]{\ensuremath{\langle{#1} \rangle}}

\newcommand{\Cov}{{\rm Cov}}
\renewcommand{\vec}[1]{\boldsymbol{#1}}

\newcommand{\diag}{{\rm diag}}

\newcommand{\id}{\mathbb{1}}

\newcommand{\Covn}[1]{{\rm Cov}_{\varrho_{#1}}(\boldsymbol{g}_{#1})}

\newcommand{\C}{X_\varrho}

\newcommand{\Corr}{\mathfrak X_\varrho}
\newcommand{\CorrA}{\mathfrak A_\varrho}
\newcommand{\CorrB}{\mathfrak B_\varrho}
\newcommand{\Corand}{\mathfrak R_\varrho}
\newcommand{\Cored}{\mathfrak X_\varrho^{\mathrm{su(d)}}}
\newcommand{\CoredA}{\mathfrak A_\varrho^{\mathrm{su(d)}}}
\newcommand{\CoredB}{\mathfrak B_\varrho^{\mathrm{su(d)}}}
\newcommand{\Ca}{v_a}
\newcommand{\Cb}{v_b}
\newcommand{\redu}{^{\mathrm{su(d)}}}

\newcommand{\idmap}{{\rm id}} 

\newtheorem{observation}{Observation}

\usepackage{cleveref}
\crefname{equation}{Eq.}{Eqs.}
\crefname{figure}{Fig.}{Figs.}
\crefname{observation}{Obs.}{Obs.}
\crefname{corollary}{Corollary}{Corollaries}
\crefname{lemma}{Lemma}{Lemmata}
\crefname{proof}{Proof}{Proofs}
\creflabelformat{proof}{#2proof#3}
\crefname{remark}{Remark}{Remarks}
\crefname{prop}{Proposition}{Propositions}

\usepackage{hyperref}
\hypersetup{citecolor=blue}
\hypersetup{colorlinks=true}
\hypersetup{linkcolor=blue}
\hypersetup{urlcolor=blue}

\begin{document}

\title{Characterizing entanglement dimensionality from randomized measurements}

\author{Shuheng Liu}
\affiliation{State Key Laboratory for Mesoscopic Physics, School of Physics, Frontiers Science Center for Nano-optoelectronics and Collaborative Innovation Center of Quantum Matter, Peking University, Beijing 100871, China}
\affiliation{Vienna Center for Quantum Science and Technology, Atominstitut, TU Wien, Vienna 1020, Austria}

\author{Qiongyi He}
\email{qiongyihe@pku.edu.cn}
\affiliation{State Key Laboratory for Mesoscopic Physics, School of Physics, Frontiers Science Center for Nano-optoelectronics and Collaborative Innovation Center of Quantum Matter, Peking University, Beijing 100871, China}
\affiliation{Collaborative Innovation Center of Extreme Optics, Shanxi University, Taiyuan, Shanxi 030006, China}
\affiliation{Peking University Yangtze Delta Institute of Optoelectronics, Nantong, Jiangsu 226010, China}
\affiliation{Hefei National Laboratory, Hefei 230088, China}

\author{Marcus Huber}
\email{marcus.huber@univie.ac.at}
\affiliation{Vienna Center for Quantum Science and Technology, Atominstitut, TU Wien, Vienna 1020, Austria}
\affiliation{Institute for Quantum Optics and Quantum Information (IQOQI), Austrian Academy of Sciences, Vienna 1090, Austria}

\author{Otfried G\"uhne}
\affiliation{Naturwissenschaftlich-Technische Fakult{\"a}t, Universit{\"a}t Siegen, Walter-Flex-Stra{\ss}e 3, Siegen D-57068, Germany}

\author{Giuseppe Vitagliano}
\email{giuseppe.vitagliano@tuwien.ac.at}
\affiliation{Vienna Center for Quantum Science and Technology, Atominstitut, TU Wien, Vienna 1020, Austria}

\begin{abstract}
    We consider the problem of detecting the dimensionality of entanglement with the use of correlations between measurements in randomized directions. First, exploiting the recently derived covariance matrix criterion for the entanglement dimensionality [S. Liu \textit{et al.}, arXiv:2208.04909], we derive an inequality that resembles well-known entanglement criteria, but contains different bounds for the different dimensionalities of entanglement. This criterion is invariant under local changes of $su(d)$ bases and can be used to find regions in the space of moments of randomized correlations, generalizing the results of [S. Imai \textit{et al.}, Phys. Rev. Lett. 126, 150501 (2021)] to the case of entanglement-dimensionality detection. In particular, we find analytical boundary curves for the different entanglement dimensionalities in the space of second- and fourth-order moments of randomized correlations for all dimensions $d_a=d_b=d$ of a bipartite system.
    We then show how our method works in practice, also considering a finite statistical sample of correlations, and we also  show that it can detect more states than other entanglement-dimensionality criteria available in the literature, thus providing a method that is both very powerful and potentially simpler in practical scenarios. We conclude by discussing the partly open problem of the implementation of our method in the multipartite scenario.
\end{abstract}

\maketitle

\section{INTRODUCTION}

In recent years an impressive degree of control over a genuinely quantum system has been reached, and 
quantum states are generated in more and more complex experiments, involving more and more particles. In these experiments it is very important to characterize entanglement, even quantitatively, as a major figure of merit for the nonclassicality of the correlations among the particles.
In particular, the so-called entanglement dimensionality plays a prominent role, as it quantifies in some sense the dimension needed for reproducing the states' correlations among a given bipartition~\cite{FriisNatPhys19}.
In fact, a high Schmidt number across many bipartitions is also the bottleneck for simulating classically 
many-body quantum states, for example through algorithms based on tensor networks~\cite{Orus2019}.
Furthermore, high-dimensional entanglement has been proven useful for several practical tasks, ranging from
improved security for quantum cryptography~\cite{ZhangQCrypto13,HuberPawlowski13}, to noise resistant quantum communication \cite{ecker19,hu20} and universal quantum computation~\cite{Lanyon2009,VandenNest13}.
A variety of methods for bounding the entanglement dimensionality has been developed and successfully implemented in several experiments over the past years~\cite{GuheneToth09,FriisNatPhys19}.
Most of these methods require quite demanding measurement capabilities, which in particular cannot always be implemented in very complex experiments, for example with many-body systems.

At the same time, a powerful idea, which has been put forward in recent years, is based on randomized measurements, which avoids the need of carefully tuning the measurement directions and bypasses even the requirement of a common reference frame between the particles.
In that approach, two parties $a$ and $b$ perform measurements of some observables $M_a$ and $M_b$, which are then randomly rotated via local unitaries $U_a\otimes U_b$. Consequently the correlations arising from such measurements are randomly distributed, and the typical theoretical approach is to investigate a sampling from the unitary Haar measure. The idea is then to estimate moments of such randomized correlations and use them to extract global information about the state.
Randomized measurements have been proven useful already for a number of tasks,
including the violation of Bell inequalities~\cite{Liang_2010,Wallman_2012,Palsson_2012,Shadbolt_2012,de_Rosier_2017}, the estimation of entanglement measures via R\'enyi entropies~\cite{vanEnk2012,Elben_2018,Brydges_2019,Elben_2019}, the estimation of fidelities~\cite{FlammiaLiu11,Elben_2020c}, topological invariants~\cite{Elben_2020a} and the detection of entanglement via negativity of the partial transpose~\cite{Zhou2020,Elben_2020b}, or directly with moments of randomized correlations~\cite{Tran_2015,Tran_2016,Dimi__2018,Saggio_2019,Ketterer_2019,Ketterer_2020,Knips_2020,Ketterer_2022}. 

While most of the works considered systems of qubits, entanglement detection for qudits of dimension higher than two has been introduced by making use of second moments of randomized correlators $\int {\rm d} U_a {\rm d} U_b \av{(U_a\otimes U_b)M\otimes M (U_a\otimes U_b)^\dagger}^2$ of certain observables $M$~\cite{Tran_2016}.
Recently, proposals to make also use of fourth-order moments have been explored~\cite{Ketterer_2019,Ketterer_2020}, including for qudits of dimension higher than two~\cite{imai2021bound}.
At this point, the question arises whether the dimensionality of entanglement can be also witnessed via randomized correlators. 
One approach that goes in that direction has been explored in the context of average work-extraction protocols~\cite{Imai_2023} and makes use of the so-called $r$-reduction criterion.
Since we want to consider measurements averaged over all local unitaries, we have to look for criteria based on local unitary invariants, such as norms of the correlations tensor~\cite{klockl2015characterizing}. 
For example, an expression that seems suitable to implement and exploit both second- and fourth-order moment of the randomized correlations is the one-norm of the correlation tensor, which is bounded for separable states~\cite{imai2021bound,de2006separability}.

In this paper, we first observe that the one-norm of the correlation tensor has different bounds for different entanglement dimensionalities, thus generalizing the criterion in Ref.~\cite{de2006separability}.  
We show that this follows as a corollary of the so-called covariance matrix criterion~\cite{guhnecova,gittsovich08,GittsovichPRA10}, which has been recently extended to detect the entanglement dimensionality~\cite{liu2022bounding}.
The condition we derive here can also be related to other well-known entanglement criteria, like the computable cross norm or realignment (CCNR) criterion~\cite{Rudolph_2005,chen03,Rudolph2003} to witness the entanglement dimensionality. 
Second, we present a practical method that uses such local unitary invariant condition to bound the entanglement dimensionality from
randomized measurements.
As a case study we show how our method works in dimension three and four and
discuss the potential limitations for increasing dimensions.
Finally, we compare our method with other prominent criteria, like witnesses given by fidelities to target entangled states, or the two-norm of the correlation tensor and discussing the practical applications with a finite sample of randomized correlations.
We conclude by summarizing our results and discussing the partly open question of extending this method to the multipartite case, i.e., to the detection of the Schmidt number across every bipartition of a multiparticle state.

\section{METHODS}

\subsection{Schmidt number and its detection}

Let us consider a bipartite system with Hilbert space $\hil = \mathbb C^{d_a} \otimes \mathbb C^{d_b}$ and a pure state $\ket{\psi} \in \hil$. 
Any arbitrary bipartite pure state can be brought to the normal form under local unitary operations $\ket{\psi} \mapsto U_a \otimes U_b \ket{\psi} = \sum_j \sqrt{\lambda_j} \ket{j_a j_b}$,
where $\{\ket{j_n}\}$ with $n=a,b$ are orthonormal basis for the two local Hilbert spaces and $\lambda_j$ are called {\it Schmidt coefficients}. 
This is termed {\it Schmidt decomposition} and can be achieved essentially from the singular value decomposition of the coefficient matrix of the state $\ket{\psi}=\sum_{kl} c_{kl}\ket{k,l}$.
In other words, the transformation that brings the coefficient matrix $c_{kl}$ in its singular value decomposition amounts to a local unitary transformation of $\ket{\psi}$, which in turn provides the Schmidt decomposition. 
The Schmidt decomposition also gives the reduced density matrix of each party as $\varrho_a = \tr_b( \ketbra{\psi} ) = \sum_j \lambda_j \ketbra{j_a}$, and similarly for $\varrho_b$.
Clearly, such a transformation does not map a product state into an entangled state nor vice versa. Moreover, it cannot change the amount of entanglement of the state, since all entanglement monotones are 
by definition invariant under local unitary transformations.
In fact, the number $s(\ket{\psi})$ of nonzero Schmidt coefficients of a pure state defines the entanglement monotone called Schmidt rank. 
This monotone can be extended to all bipartite density matrices, i.e., trace-class operators $\varrho \in \mathcal B(\hil)$, via {\it convex-roof construction}, this way defining the {\it Schmidt number}~\cite{Barbara2000SchmidtNumber}:

\be\label{eq:SNdefinition}
s(\varrho):= \inf_{\mathcal D(\varrho)} \max_{\ket{\psi_i} \in \mathcal D(\varrho)} s(\ket{\psi_i}) ,
\ee
where the infimum is taken over all pure-state decompositions $\mathcal D(\varrho)=\{p_i, \ket{\psi_i} \}$ of the density matrix $\varrho=\sum_i p_i \ketbra{\psi_i}$.
Roughly speaking, this entanglement monotone is connected with the minimal dimension of the Hilbert space needed to reproduce the states' correlations. Note that other notions can be given connected to the dimensionality of entanglement, such as that of genuine multilevel entanglement~\cite{Kraft_2018}.

Typical methods useful, especially in theory, to bound the Schmidt number from below involve either the evaluation of entanglement witnesses, i.e., hermitian operators $W_r$ such that $\tr(W_r \varrho_r)\geq 0$ for all states $\varrho_r$ such that $s(\varrho_r) \leq r$ and $\tr(W_r \varrho)< 0$ for some state $\varrho$ such that $s(\varrho) > r$, or the application of so-called {$r$-positive maps}, i.e., linear maps $\mathcal L_r: \mathbb C^{d_a} \otimes \mathbb C^{d_a} \rightarrow \mathbb C^{d_b} \otimes \mathbb C^{d_b}$ such that $\idmap_r \otimes \mathcal L_r$ is positive, where $\idmap_r$ is the identity map in the space of $r\times r$ matrices~\cite{Barbara2000SchmidtNumber,Sanpera2001SchmidtNumber,Huber_2018}.

The most typical Schmidt-number witnesses are operators of the type 
\be\label{eq:fidelityWit}
W_r=\sum_{k=1}^r\lambda_{k}\id - \ketbra{\Psi_r} ,
\ee
where $\ket{\Psi_r}=\sum_{k} \sqrt{\lambda_k} \ket{k k}$ is a given state with Schmidt rank larger than or equal to $r$ and Schmidt coefficients ordered decreasingly~\cite{FlammiaLiu11,Krenn6243Generation,Erker2017Quantifying,BavarescoNatPhys18,Sanpera2001SchmidtNumber}.
This means that the fidelity with a given target state is always upper bounded by its $r$ greatest Schmidt coefficients for all states with the same Schmidt number, i.e., 
\be\label{eq:fidelityCrit}
\tr(\varrho \ketbra{\Psi_r}) \leq \sum_{k=1}^r\lambda_{k}
\ee
must hold for all states $\varrho$ such that $s(\varrho) \leq r$.
In particular, for a $d$-dimensional maximally entangled state the upper bound on the fidelity as in \cref{eq:fidelityCrit} is given by $r/d$, since all Schmidt coefficients are equal to each other. 

The most typical $r$-positive map used for Schmidt-number detection is given by a generalization of the {\it reduction map}~\cite{reductionCritH,reductionCritC}, namely $\mathcal D_r (A) = \tr(A) \id - \tfrac 1 r A$. 
When this map is applied locally on a bipartite density matrix one gets
\be\label{eq:rRedCrit}
\idmap_a \otimes \mathcal D_r (\varrho ) = \varrho_a \otimes \id^{(b)} - \tfrac 1 r \varrho \geq 0 ,
\ee
and its positivity must hold for all states with Schmidt number smaller than $r$~\cite{Barbara2000SchmidtNumber}.
Note that this map is dual to the witness with respect to the maximally entangled state, namely, the witness in \cref{eq:fidelityWit} with $\ket{\Psi_r}$ being the $r$-dimensional maximally entangled state is obtained from the map
$\idmap_a \otimes \mathcal D_r$ after applying the Choi-Jamio\l kowski isomorphism. 

A further important Schmidt-number criterion can be found from the coefficients of expansion of the density matrix in a pair of local bases, namely we have $\varrho = \sum_{kl} (\Corr)_{kl} g_k^{(a)} \otimes g_l^{(b)}$.
In particular, considering local bases that are composed of $\id/\sqrt{d}$ plus normalized $su(d)$ bases and defining the submatrix $\Cored$ by omitting the entries $\av{\tfrac{\id^{(a)}}{\sqrt d} \otimes g_{j}^{(b)}}$ and $\av{ g_{i}^{(a)} \otimes \tfrac{\id^{(b)}}{\sqrt d}}$, it is possible to find a criterion for the entanglement dimensionality that reads~\cite{klockl2015characterizing}
\begin{equation}\label{eq:HuberCrit}
\|\Cored\|_2^2 = \sum_{kl} (\Cored)_{kl}^2 \leq 1+\frac{r-2d}{d^2 r} . 
\end{equation}
 
Finally, a more general approach can be followed by starting from the (symmetric) covariance matrix associated to a pair of $\obsset{g} = (\obsset{g}_a, \obsset{g}_b)$, where  $\obsset{g}_a=(g_1^{(a)} \otimes \id , \dots , g_{d_a^2}^{(a)} \otimes \id)$ and 
$\obsset{g}_b=(\id \otimes g_1^{(b)} , \dots , \id \otimes g_{d_b^2}^{(b)})$ are orthonormal basis of observables for party $a$ and $b$, respectively.
For a general set of observables $\obsset M=(M_1,\dots,M_K)$ and a given state $\varrho$, the (symmetric) covariance matrix has components given by
\be 
[\Cov_\varrho(\obsset M)]_{jk} = \tfrac 1 2 \av{M_j M_k + M_k M_j}_\varrho - \av{M_j}_\varrho \av{M_k}_{\varrho} ,
\ee
and for a pair of local operator bases it assumes the block form
\begin{equation}\label{eq:blockCov}
\Cov_\varrho(\obsset{g}) := \Gamma_\varrho = 
\left(\begin{array}{ll}
\gamma_a & \C \\
\C^T & \gamma_b
\end{array}\right),
\end{equation}
in which the diagonals $\gamma_a:=\Covn{a}$ and $\gamma_b:=\Covn{b}$ are the symmetric covariance matrices of each party, and the off-diagonal blocks
\be
\begin{aligned}
(\C)_{kl} &= \av{g_k^{(a)} \otimes g_l^{(b)}}_\varrho - \av{g_k^{(a)}}_\varrho \av{g_l^{(b)}}_\varrho 
&:= (\Corr)_{kl} - (\Ca)_k (\Cb)_l
\end{aligned}
\ee
are the cross covariances between the two local observables vectors, and we define the marginals $\varrho_a=\tr_{b}(\varrho)=\sum_k (\Ca)_k g_k^{(a)}$ and 
$\varrho_b=\tr_{a}(\varrho)=\sum_k (\Cb)_k g_k^{(b)}$. Note that we have $(\Ca)_k = \sum_l (\Corr)_{kl} \tr(g_l^{(b)})$ and similarly for party $b$.

Based on this covariance matrix, a compact criterion that has to hold for all states with a certain Schmidt number $r$ can be found~\cite{liu2022bounding} that reads
\be\label{eq:CMCrankr}
\Gamma_{\varrho} - \sum_k p_k \Gamma^{(k)}_r \geq 0 ,
\ee
where (omitting the label $k$ for simplicity) $\Gamma_r=\Cov_{\psi_r}(\obsset{g})$ are (positive) covariance matrices associated to pure Schmidt-rank-$r$ states $\psi_r=\sum_j \sqrt{\lambda_j}\ket{j_a j_b}$
and $\{p_k\}$ is some probability distribution. 
Here, each $\Gamma_r$ has in general a block form
\be\label{eq:Gammarblock}
\Gamma_r =
\left(\begin{array}{ll}
\kappa_a & X_r \\
X_r^T & \kappa_b
\end{array}\right),
\ee
with the singular values of the blocks satisfying
\be 
\begin{aligned}\label{eq:kappaCconds}
\vec{\epsilon}(\kappa_a) &= \vec \epsilon(\kappa_b) = (\vec{\epsilon}(D),\{\tfrac 1 2 (\lambda_j+\lambda_k)\}_{j<k=1}, \{\tfrac 1 2 (\lambda_j+\lambda_k)\}_{j>k=1} ) , \\
\vec{\epsilon}(|X_r|) &=(\vec{\epsilon}(D),\{\sqrt{\lambda_j \lambda_k}\}_{j<k=1},\{\sqrt{\lambda_j \lambda_k}\}_{j>k=1}),
\end{aligned}
\ee 
where $D$ is the $d_a\times d_b$ matrix with elements $D_{jk}= \lambda_j \delta_{jk}-\lambda_j \lambda_k$.

\subsection{Haar-randomized correlators and Bloch-sphere integrals}

One potential advantage of criteria based on quantities, which are invariant under change of local $su(d)$ bases, is that they can be  evaluated from random measurements, if specifically sampled from the local unitary Haar measures. To see that let us consider correlations between generic observables $M_a$ and $M_b$, randomized over local unitaries, namely
$\av{U_a M_a U_a^\dagger \otimes U_b M_b U_b^\dagger}_\varrho$, 
where $U_a \otimes U_b$ is a local unitary randomly sampled from the Haar measure.
The probability distribution of the correlations arising from such randomized measurements can be characterized by its moments
\begin{equation}\label{eq:Emoments}
\Corand^{(m)}(M_a \otimes M_b)=\int {\rm d} U_a {\rm d} U_b\left[\av{U_a M_a U_a^\dagger \otimes U_b M_b U_b^\dagger}_\varrho\right]^m ,
\end{equation}
which are evaluated from the Haar measures ${\rm d} U_a$ and ${\rm d} U_b$.

The nice feature of such Haar-randomized correlators is that these moments written above in some cases can be calculated 
by transforming the integration over the unitary Haar measure into an integration over $d$-dimensional Bloch spheres, namely
\be\label{eq:BLSPint}
\mathcal{S}_\varrho^{(m)} := N \int {\rm d} \hat{\vec \alpha} \int {\rm d} \hat{\vec \beta} \left[\av{\sigma_{\hat{\vec \alpha}}\otimes \sigma_{\hat{\vec \beta}}}_\varrho \right]^m ,
\ee
where $\sigma_{\hat{\vec \alpha}} := \hat{\vec \alpha} \cdot \vec \sigma$ and $\sigma_{\hat{\vec \beta}} := \hat{\vec \beta} \cdot \vec \sigma$ with $\vec \sigma = (\sigma_1, \dots, \sigma_{d^2-1})$ being an $su(d)$ basis and $(\hat{\vec \alpha},\hat{\vec \beta})$ being two real $(d^2-1)$-dimensional unit vectors. Furthermore, $N$ is a normalization that is chosen such that 
$\mathcal{S}_\varrho^{(m)}=1$ for pure product states.
One important property that makes this possible is that both $\mathcal{S}_\varrho^{(m)}$ and $\Corand^{(m)}(M_a \otimes M_b)$ as defined above are invariant under local unitaries. 
However, the value of the latter depends on the choice of the observables $M_a$ and $M_b$, in particular on their concrete eigenvalues (not on the eigenvectors due to invariance under unitaries).  
On the other hand, the $\mathcal{S}_\varrho^{(m)}$ are actually invariant over all changes of local $su(d)$ bases, which results in the fact that they depend only on the singular values of the correlation matrix $\Cored$.

In fact, after some algebra it has been shown in Ref.~\cite{imai2021bound} that one obtains\footnote{Note that with respect to Ref.~\cite{imai2021bound} we use a different normalization.} 
\begin{subequations}\label{eq:ConstS}
\begin{eqnarray}
&\mathcal{S}_\varrho^{(2)}=& \frac{d^2}{(d-1)^2} \sum_{i=1}^{d^2-1} \epsilon_i^2 , \label{eq:ConstS2}  \\
&\mathcal{S}_\varrho^{(4)}=& \frac{2d^4}{3(d-1)^4}  \sum_{i=1}^{d^2-1} \epsilon_i^4+\frac{1}{3}\left(\mathcal{S}_\varrho^{(2)}\right)^2 , \label{eq:ConstS4}
\end{eqnarray}
\end{subequations}
where $\epsilon_i:=\epsilon_i(|\Cored|)$ are singular values of $\Cored$. See \cref{app:selfcontained} for details.

Nevertheless, Ref.~\cite{imai2021bound} has also shown that  by choosing appropriately the eigenvalues of $M=M_a=M_b$ it is possible to obtain $(d+1)^2\Corand^{(2)}(M \otimes M)=\mathcal{S}_\varrho^{(2)}$ and $\tfrac{(d+1)^2\left(d^2+1\right)^2}{9(d-1)^2}\Corand^{(4)}(M \otimes M) = \mathcal{S}_\varrho^{(4)}$ (see also Ref.~\cite{nikolai}). In particular, the observable $M$ has the form $M=\diag(\alpha_+,\dots,\alpha_+,\beta,\alpha_-,\dots,\alpha_-)$ for certain real numbers $\alpha_\pm$ and $\beta$ that depend on the dimension.~\cite{imai2021bound}

\section{MAIN RESULTS}

\subsection{New criterion based on Bloch invariants}

Once the relation between moments of randomized correlations $\Corand^{(m)}$ and expressions like \cref{eq:ConstS} have been found, one might think to use the former measurements to witness important properties of the quantum state, e.g., entanglement~\cite{imai2021bound}. For this task one needs a further ingredient, namely a suitable entanglement criterion written in terms of the local orthogonal invariants $\epsilon_i$. 
Similarly, finding a witness of the entanglement dimensionality in terms of the $\epsilon_i$ would enable the possibility of detecting the former with randomized correlators via similar methods.

In the following, we show that a condition like that can be derived from the covariance matrix criterion in \cref{eq:CMCrankr}.

\begin{observation}\label{cor:BlochRep}
For every Schmidt-number-r density matrix $\varrho$
the following inequality holds
\begin{equation}\label{eq:CoredCrit}
\tr|\Cored|- (r-1) \leq \sqrt{\left(1-\frac{1}{d_{a}}\right)\left(1-\frac{1}{d_{b}}\right)} ,
\end{equation}
which is also invariant under change of orthonormal pairs of $su(d_a)$ and $su(d_b)$ basis. 
Here trace norm is defined as $\tr|A|=\tr\sqrt{A^\dagger A}$.
\end{observation}

{\it Proof.---}We can rewrite \cref{eq:CMCrankr} in the form
\begin{equation}
\underbrace{\left(\begin{array}{cc}
\CorrA-\kappa_{a} & \Corr-X_r \\
\Corr^{T}-X_r^{T} & \CorrB-\kappa_{b}
\end{array}\right)}_{Y}-\left(\begin{array}{l}
\Ca \\
\Cb
\end{array}\right)\left(\begin{array}{l}
\Ca \\
\Cb
\end{array}\right)^{T} \geq 0 ,
\end{equation}
where we called $\mathfrak{A} := \gamma_a + \Ca \Ca^T$ and $\mathfrak{B}:= \gamma_b + \Cb \Cb^T$ the single-particle correlation matrices.
Since the term we subtract is a projector, $Y$ itself is positive and so is its principal minor 
\begin{equation}
\left(\begin{array}{cc}\CoredA-\kappa_{a}\redu & \Cored-X_r\redu \\ (\Cored-X_r\redu)^T & \CoredB-\kappa_{b}\redu\end{array}\right)\geq 0 ,
\end{equation}
where we consider two local bases composed of identity plus $su(d)$ bases and have eliminated the first row and first column from each block.
In other words, the above matrix contains only variances and cross correlations relative to the $su(d_a)$ and $su(d_b)$ bases for the two particles.

Multiplying this matrix $Y$ with vectors $\vec t_\mu=\left(\alpha \vec u_\mu, -\beta \vec v_\mu \right)$, where $\vec u_\mu$ and $\vec v_\mu$ are the left and right singular vectors of $\Corr$ and $\alpha,\beta>0$ and summing over all $\mu$ we obtain $\alpha^2 \tr(\CoredA) + \beta^2 \tr(\CoredB) -2\alpha\beta \tr|\Cored| 
= \alpha^2 (d_a-\tfrac 1 {d_a}) + \beta^2 (d_b-\tfrac 1 {d_b}) -2\alpha\beta \tr|\Cored| 
\geq \alpha^2 \tr(\kappa_{a}\redu) + \beta^2 \tr(\kappa_{b}\redu) -2\alpha\beta \sum_\mu |(X_r\redu)_{\mu \mu}| 
\geq \alpha^2 \tr(\kappa_{a}\redu) + \beta^2 \tr(\kappa_{b}\redu) -2\alpha\beta \tr|X_r\redu| 
\geq \alpha^2 [d_a - 1 + E_L(\psi_r)] + \beta^2 [d_b - 1 + E_L(\psi_r)] -2\alpha\beta [r- 1 + E_L(\psi_r)]$,
where $E_L(\psi_r) = 1 - \tr[(\tr_a \kb{\psi_r}{\psi_r} )^2]$ is the linear entropy of the single-particle reduced density matrix of a generic (optimal) pure Schmidt rank-$r$ state (which is the same for both parties). Here we use the relations $\tr(\CoredA) = d_a - \tfrac 1 {d_a}$, $\tr(\CoredB) = d_b-\tfrac 1 {d_b}$, $\tr(\kappa_{s}\redu) = [d_s - 1 + E_L(\psi_r)]$ for $s=a,b$ and $\tr|X_r\redu|\leq \tr|X_r|\leq r- 1 + E_L(\psi_r)$. The latter can be directly verified using the generic expression of its singular values from \cref{eq:kappaCconds}. This way, we obtain $(\alpha - \beta)^2 \geq (\alpha - \beta)^2 [1-E_L(\psi_r)] \geq \tfrac{\alpha^2}{d_a}+\tfrac{\beta^2}{d_b} + 2\alpha\beta (\tr|\Cored|-r)$. 
Now, minimizing over all $\alpha$ and $\beta$ we have that the minimum is achieved for $\alpha/\beta=(\tr|\Cored|-r+1)/(1-\tfrac 1 {d_a})$, whenever this is positive and results in \cref{eq:CoredCrit}. 
This criterion is invariant under changes of orthonormal pairs of $su(d_a)$ and $su(d_b)$ bases, since any such change of bases results in local orthogonal transformations of the correlation matrix, namely $\Cored \mapsto O_a \Cored O_b^T$, under which the sum of singular values is invariant. 
\qed

In the following we restrict to bipartite spaces with equal dimensions $d_a=d_b=d$, in order to discuss the implementation of this criterion in a method to witness entanglement dimensionality from the randomized correlators $\Corand^{(2)}$ and $\Corand^{(4)}$. For equal dimensions, \cref{eq:CoredCrit} reduces to 
\begin{equation}\label{eq:CritCoredD}
\tr|\Cored| := \sum_{k=1}^{d^2-1} \epsilon_{k} \leq r-\frac{1}{d} .
\end{equation}
Note that for $r=1$ the left-hand side is exactly the same as in the entanglement criterion from Ref.~\cite{de2006separability}, which was also used in Ref.~\cite{imai2021bound} to find entanglement witnesses from randomized the correlators $\Corand^{(2)}(M \otimes M)$ and $\Corand^{(4)}(M \otimes M)$.
Thus, our condition can be seen as a generalization of such a criterion to the entanglement dimensionality.
Moreover, our condition can be compared to \cref{eq:HuberCrit}, since they are based on basically the same information. 
The only difference is that \cref{eq:HuberCrit} considers a different matrix norm on the left-hand side.

Note also that expressing $\Corr$ in its singular value decomposition $\Corr = O_a^T \cdot \xi \cdot O_b$ corresponds to expressing the density matrix as $\varrho = \sum_k \xi_{k=0}^{d^2-1} G_k^{(a)} \otimes G_k^{(b)}$ with $G_k^{(a)} = \sum_l (O_a)_{kl} g_l^{(a)}$ obtained from the initial canonical bases $g_l^{(a)}$ and with $\xi_k\geq 0$. This is a Schmidt decomposition in operator space, with $\obsset G_a$ and $\obsset G_b$ being the Schmidt vectors in operator space. Now, expressing the covariance matrix in these operator Schmidt bases we can also derive the following Schmidt-number-$r$ necessary condition:
\be\label{eq:epsiloncrit}
\sum_{k=0}^{d^2-1} \xi_k \leq r ,
\ee
which in turn reduces to \cref{eq:CritCoredD} in the case in which $\obsset G_a$ and $\obsset G_b$ are of the form identity plus an $su(d)$ basis. At the same time, it is worth emphasizing that in general the singular value of the full correlations matrix, i.e., the $\xi_k$ do not necessarily coincide with the singular values of the $su(d)$ submatrix, i.e., the $\epsilon_k$.

The condition \eqref{eq:epsiloncrit} can be shown from the following corollary of \cref{eq:CMCrankr}, which was presented in Ref.~\cite{liu2022bounding}:
\be\label{eq:Cor2PRL}
\tr|\C|- (r-1) \leq
\sqrt{[1 - \tr(\varrho^2_a)][1 -\tr(\varrho^2_b)]} ,
\ee
which is invariant under a full local orthogonal change of bases. 

{\it Proof of \cref{eq:epsiloncrit}.---} We have the following chain of inequalities~$\sum_k~\xi_k~-~\sum_k~\xi^2_k~|\tr(G_k^{(a)}) \tr(G_k^{(b)})|~-(r-1)\leq \sum_k \left|\xi_k - \xi^2_k \tr(G_k^{(a)}) \tr(G_k^{(b)}) \right|- (r-1)\leq \tfrac 1 2 [ 2 - \tr(\varrho^2_a) -\tr(\varrho^2_b) ]= ( 1 - \tfrac 1 2 \sum_k \xi^2_k [(\tr(G_k^{(a)}))^2 + (\tr(G_k^{(b)}))^2 ])\leq 1 - \sum_k \xi^2_k |\tr(G_k^{(a)}) \tr(G_k^{(b)})|$. Here, the first is due to $|a-b|\geq |a|-|b|$, the second is obtained by substituting the operators Schmidt decomposition into \cref{eq:Cor2PRL} and using $\sqrt{ab}\leq \tfrac 1 2 (a+b)$. Finally, the last inequality is similarly obtained from~$a^2+b^2\geq~2~|ab|$. 
\qed

Observe also how the expression in \cref{eq:epsiloncrit} is similar to the CCNR criterion, which can be put in the form $\sum_k \xi_k \leq 1$~\cite{Rudolph_2005,chen03,Rudolph2003}, and in a sense generalizes it to detect different Schmidt numbers.
Such a generalization has been already derived in Ref.~\cite{JohnstonKribs15}, where the authors actually derived an even stronger criterion based on unitarily invariant norms of the density matrix.

\subsection{Entanglement dimensionality witness from randomized measurements}

With all the ingredients discussed in the previous sections we are now in the position of 
generalizing the method of Ref.~\cite{imai2021bound} to witness the entanglement dimensionality.
Thus, we consider the moments $\Corand^{(2)}(M \otimes M)$ and $\Corand^{(4)}(M \otimes M)$ for the specific $d$-dimensional observable $M$
that is such that $\Corand^{(2)}(M \otimes M) \propto \mathcal{S}_\varrho^{(2)}$ and $\Corand^{(4)}(M \otimes M)\propto \mathcal{S}_\varrho^{(4)}$.
Then, varying the $\epsilon_i$ and making use of the criterion in \cref{eq:CritCoredD}
we can calculate the boundary regions of Schmidt-number-$r$ states in the plane $(\mathcal{S}_\varrho^{(2)},\mathcal{S}_\varrho^{(4)})$ by solving the problem
\begin{equation}\label{eq:SchmidtProb}
\begin{aligned}
\max_{\epsilon_i} / \min_{\epsilon_i} \quad  & \mathcal{S}_\varrho^{(4)}(\epsilon_i)  \\
\text { subject to } \quad  & \mathcal{S}_\varrho^{(2)}=\tfrac{d^2}{(d-1)^2} \sum_{i=1}^{d^2-1} \epsilon_i^2, \\
& \sum_{i=1}^{d^2-1} \epsilon_i \leq r-\tfrac 1 d, \\
& \epsilon_i \geq 0 , 
\end{aligned}
\end{equation}
namely by maximizing and minimizing the value of $\mathcal{S}_\varrho^{(4)}$ for each fixed value of $\mathcal{S}_\varrho^{(2)}$
with the additional constraints given by \cref{eq:CritCoredD,eq:ConstS} (plus positivity of the $\epsilon_i$).

\begin{figure}[t]
\centering
\includegraphics[width=0.48\textwidth]{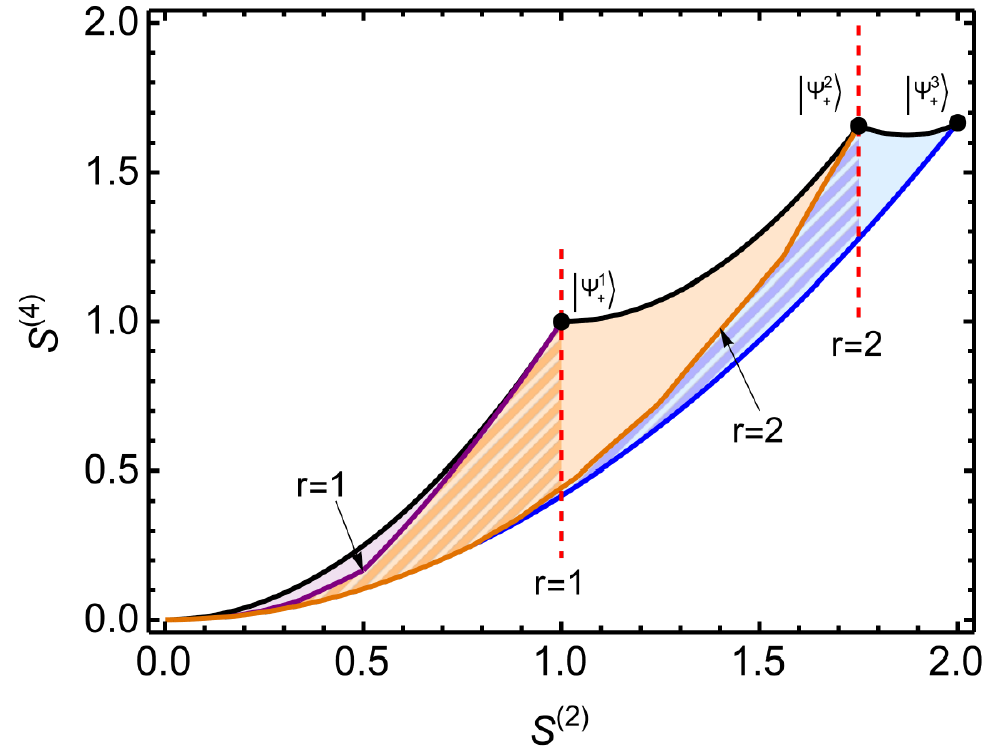}
\caption{Boundary-region plane for the different Schmidt numbers in the $(\mathcal{S}_\varrho^{(2)},\mathcal{S}_\varrho^{(4)})$ in $d=3$ as coming from solving \cref{eq:SchmidtProb}. Solid lines are as follows: upper bound for all states and lower bounds for $r=1,2,3$, respectively. Dashed vertical lines correspond to the criterion in \cref{eq:HuberCrit} for the different Schmidt numbers.
}
\label{fig:CriterionBoundariesD3}
\end{figure}

\begin{figure}[t]
\centering
\includegraphics[width=0.48\textwidth]{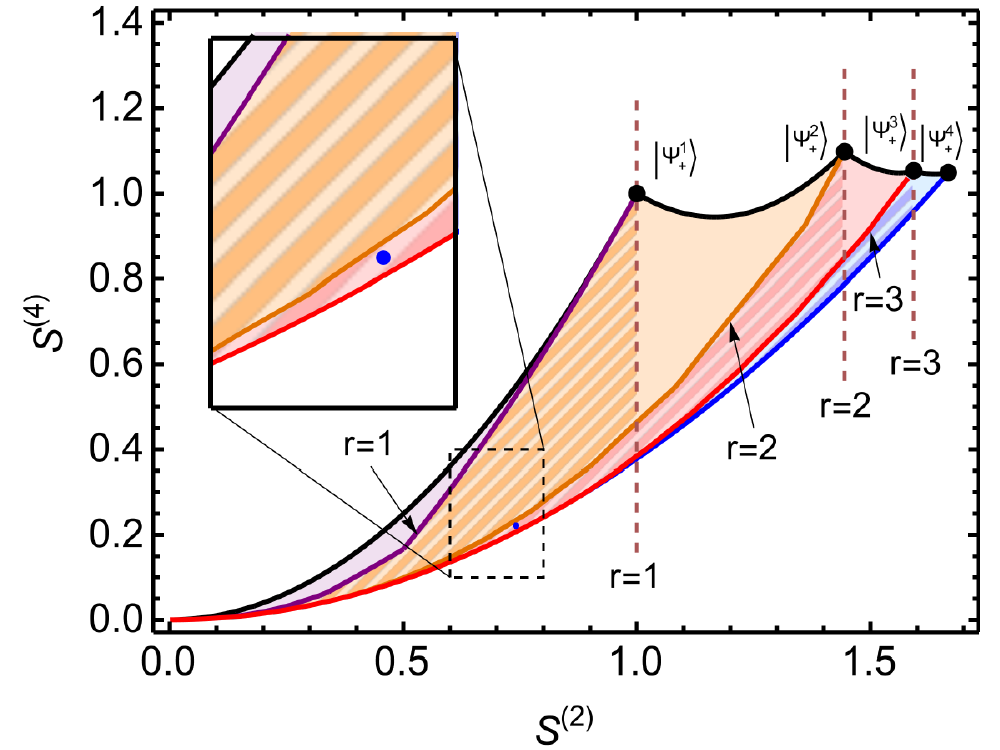}
\caption{Boundary-region plane for the different Schmidt numbers in the $(\mathcal{S}^{(2)},\mathcal{S}^{(4)})$ in $d=4$ as coming from solving \cref{eq:SchmidtProb}. Solid lines are: upper bound for all states and lower bounds for $r=1,2,3,4$ respectively. Dashed vertical lines correspond to the criterion in \cref{eq:HuberCrit} for the different Schmidt numbers.
The blue dot corresponds to the Schmidt-number-$3$ state $\varrho_W$ as defined in the main text. 
}
\label{fig:CriterionBoundariesD4}
\end{figure}

Such boundaries for $d=3,4$ and the different Schmidt numbers are plotted in \cref{fig:CriterionBoundariesD3,fig:CriterionBoundariesD4}. It can be observed numerically that the depicted regions are fully filled with physical states (cf. \cref{fig:d3randomsamples} in \cref{app:boundarystates}). Extremal points are product or maximally entangled states of a given Schmidt rank $r$, which we label $\ket{\Psi_+^r} = \tfrac 1 {\sqrt{r}} \sum_{j=0}^{r-1} \ket{jj}$. 
In particular, the curves that distinguish the different Schmidt numbers are obtained from the minimization problem. These curves can be obtained analytically for all dimensions with the method of Lagrange multipliers and the Karush-Kuhn-Tucker conditions (cf. \cref{app:KKTsolutions}).
The result is, for each Schmidt number $r$, a piecewise function with pieces labeled by an integer $n=1, \cdots, d^2-2$:
\begin{equation}\label{eq:boundarycurves}
\mathcal{S}_r^{(4)}= \begin{cases}
f_0(\mathcal{S}^{(2)}) , & 0 \leq \mathcal{S}^{(2)} \leq \frac{B_r^2 }{(d^2-1)} \\ 
f_n(\mathcal{S}^{(2)}) , & \frac{B_r^2 }{(n+1)} \leq \mathcal{S}^{(2)} \leq \frac{B_r^2 }{n} 
\end{cases}
\end{equation}
where $B_r=\tfrac{d r-1}{d-1}$ and the pieces are $f_0(X):=\tfrac{d^2+1}{3\left(d^2-1\right)}X^2$ and $f_n(X)=\tfrac{2}{3(n+1)^4} \left[\left(\sqrt{g_n(X)}-B_r\right)^4+\tfrac{1}{n^3} \left(\sqrt{g_n(X)}+ n B_r\right)^4\right]+\tfrac{X^2}{3}$ with $g_n(X)=n(n+1) X-n B_r^2$.

\subsection{Detection with finite statistics and comparison with other criteria}

Here we compare our newly derived entanglement dimensionality witnesses with other Schmidt-number criteria.
Preliminarly, let us notice that the criterion \cref{eq:epsiloncrit}, detects strictly more states than any criterion constructed from the fidelity with respect to a maximally entangled state, namely $\bra{\Psi^d_+}\varrho \ket{\Psi^d_+}\leq \tfrac r d$ where $\ket{\Psi^d_+}=\sum_{j=1}^d \tfrac 1 {\sqrt{d}} \ket{j_aj_b}$.
In fact, the latter criterion can also be written as $\tr \Corr \leq r$ in some canonical bases, and it is clear that $\tr \Corr \leq \sum_k \xi_k$ for any canonical pair of local bases. 
Furthermore, there are states that can be detected with a strictly higher Schmidt number with \cref{eq:epsiloncrit}. As an example consider the mixed state $\varrho_W=\tfrac{1}{2}\kb{\Psi_+^{3}}{\Psi_+^{3}}+\tfrac{1}{4}(\ket{23}+\ket{32})(\bra{23}+\bra{32})$,  which has Schmidt number $s(\varrho_W)=3$. Its Schmidt number certified by any fidelity witness (not just with respect to maximally entangled targets) is at most $2$~\cite{WeilenmannPRL2020BeyondFidelity}, while \cref{eq:epsiloncrit}
detects the actual Schmidt number $3$. 
More in general, one can make statistics with random mixed states and observe that typically corollaries of the covariance matrix criterion like \cref{eq:Cor2PRL} detect significantly more states than any fidelity witness \eqref{eq:fidelityCrit}. The same holds for the criterion based on the $r$-reduction map in \cref{eq:rRedCrit}.

At the same time, we can observe that states like $\varrho_W$, i.e., states that are not detected by any fidelity witness are ideally detected with the correct Schmidt number by our criterion based on randomized measurements. See, for example, Fig.~\ref{fig:CriterionBoundariesD4}. Similarly, we can compare our criterion with \cref{eq:HuberCrit}. The latter can be simply expressed as $\sum_{i=1}^{d^2-1} \epsilon_i^2 = \mathcal{S}_\varrho^{(2)} \leq  1+\tfrac{r-2d}{d^2 r}$
and thus corresponds to just vertical lines in the $(\mathcal{S}^{(2)},\mathcal{S}^{(4)})$ plane, placing per s\'e no constraints on $\mathcal{S}^{(4)}$.
It seems thus reasonable to expect that generically our method would detect more states, given that both of them are saturated by some pure Schmidt-rank-$r$ states.
In fact, from \cref{fig:CriterionBoundariesD3,fig:CriterionBoundariesD4} we can observe that this is the case in $d=3,4$, where there are regions of states (depicted as striped) with Schmidt number larger than what is detected by \cref{eq:HuberCrit}. From the analytic expression of the boundary curves \cref{eq:boundarycurves} we can observe that the same happens actually for every dimension: 
both criteria provide boundaries that start from the points given by the states $\ket{\Psi_+^{r}}$. 

However, our criterion gives a tighter boundary in the rest of the plane. For example, our criterion has higher tolerance to white noise than \cref{eq:HuberCrit} for detecting the maximally entangled state, i.e., the density matrix $\varrho= (1-p)\ketbra{\Psi_+^{d}}+\frac{p}{d^2}\id_{d}\otimes \id_d$ is detected to have a given Schmidt number up to a higher $p$. In fact, we have $\tr|\Cored|=(d^2-1)\frac{1-p}{d}$ and $\|\Cored\|_2=\sqrt{d^2-1}\frac{1-p}{d}$. Substituting this into \cref{eq:CritCoredD} and \cref{eq:HuberCrit}, we can observe that our criterion gives higher upper bound to $p$. Notice also that one could use the well-known relation $\sqrt{\sum_{i=1}^x|c_i|^2}\leq\sum_{i=1}^x|c_i|\leq \sqrt{x}\sqrt{\sum_{i=1}^x|c_i|^2} $ to derive a bound on $\tr|\Cored|$ from the known bound on $\|\Cored\|_2$, which still turns out to be strictly weaker.

So far we have discussed what states are detected based on the exact calculations of the quantities $(\mathcal{S}_\varrho^{(2)},\mathcal{S}_\varrho^{(4)})$, which in principle would correspond to the moments of the randomized correlators, obtained with infinite statistics. Let us now discuss what happens when the randomized moments are evaluated from finite statistics. 
After a finite sample of $N_{\rm tot}$ random local unitaries, we can estimate the Bloch sphere integrals as
\be
\begin{aligned}
\hat{\mathcal{S}}^{(2)}_\varrho &= (d+1)^2 \tfrac{1}{N_{\rm tot}} \sum_{i} x^2_i(U_a \otimes U_b) , \\ 
\hat{\mathcal{S}}^{(4)}_\varrho &= \tfrac{(d+1)^2\left(d^2+1\right)^2}{9(d-1)^2} \tfrac{1}{N_{\rm tot}} \sum_{i}x^4_i(U_a \otimes U_b) ,
\end{aligned}
\ee
where we indicate $x_i(U_a \otimes U_b):=\av{(U_a \otimes U_b)_i M \otimes M (U_a \otimes U_b)^\dagger_i}_\varrho$.
Denoting with $\mu(\cdot)$ and $\sigma(\cdot)$ the mean and standard deviation of a random variable we have
\begin{equation}
\begin{aligned}
\mu\left(\hat{\mathcal{S}}_\varrho^{(2)}\right) & =(d+1)^2 \Corand^{(2)} , \\
\sigma^2\left(\hat{\mathcal{S}}_\varrho^{(2)}\right) & =\tfrac{(d+1)^4}{N_{\rm tot}} \left( \Corand^{(4)} - (\Corand^{(2)})^2 \right) , \\
\mu\left(\hat{\mathcal{S}}_\varrho^{(4)}\right) &=\frac{(d+1)^2\left(d^2+1\right)^2}{9(d-1)^2} \Corand^{(4)} , \\
\sigma^2\left(\hat{\mathcal{S}}_\varrho^{(4)}\right) & =\tfrac{(d+1)^4\left(d^2+1\right)^4}{81(d-1)^4 N_{\rm tot}} \left(\Corand^{(8)} - (\Corand^{(4)})^2 \right) ,
\end{aligned}
\end{equation}
which holds whenever all the individual events are independent and distributed according to the Haar measure.

Thus, with these relations, we can evaluate with how much confidence a state is detected assuming that the individual trials are independently drawn from the Haar measure.
As a paradigmatic test, we consider $d=3$, and $\ket{\Psi_{+}^3}$ as the target measured state. For this state, the ideal second- and fourth-order integrals are $\mathcal{S}_{\Psi_+^3}^{(2)}=2$ and $\mathcal{S}_{\Psi_+^3}^{(4)}=\frac{5}{3}$. We take a random numerical sample of $N_{\rm tot}=\{10^3,10^4,10^5\}$ local unitaries from the Haar measure, respectively, and obtain as the estimate for the moments of the randomized correlators the values
\begin{equation}
\begin{aligned}
\mu\left(\hat{\mathcal{S}}_{\Psi_+^3}^{(2)}\right) & =2 , \\
\sigma^2\left(\hat{\mathcal{S}}_{\Psi_+^3}^{(2)}\right) & =\tfrac 1 {N_{\rm tot}} \tfrac{28}{5} , \\
\mu\left(\hat{\mathcal{S}}_{\Psi_+^3}^{(4)}\right) & =\tfrac{5}{3} , \\
\sigma^2\left(\hat{\mathcal{S}}_{\Psi_+^3}^{(4)}\right) & \approx \tfrac{1}{N_{\rm tot}} 12.8794 .
\end{aligned}
\end{equation}
When $N_{\rm tot}=10^3,10^4,10^5$, respectively, taking three-$\sigma$ intervals as a measure of error we obtain
\begin{equation}
\begin{array}{lll}
\hat{\mathcal{S}}_{\Psi_+^3}^{(2)}=2 \pm 0.224, & \hat{\mathcal{S}}_{\Psi_+^3}^{(4)}=1.667 \pm 0.340, & N_{\rm tot}=10^3 , \\
\hat{\mathcal{S}}_{\Psi_+^3}^{(2)}=2 \pm 0.071, & \hat{\mathcal{S}}_{\Psi_+^3}^{(4)}=1.667 \pm 0.108, & N_{\rm tot}=10^4 , \\
\hat{\mathcal{S}}_{\Psi_+^3}^{(2)}=2 \pm 0.022, & \hat{\mathcal{S}}_{\Psi_+^3}^{(4)}=1.667 \pm 0.034, & N_{\rm tot}=10^5,
\end{array}
\end{equation}
and we can see that the state is detected within the error already with the sample of $N_{\rm tot}=10^3$ points, as detailed in \cref{app:finitestat}.

Similarly, one can calculate the white-noise tolerance of the method, for example with a sample of $N_{\rm tot}=10^4$ random unitaries. In this case we get that the state 
$\varrho_p := (1-p)\ketbra{\Psi_{+}^3}+\tfrac{p} 9 \id_3 \otimes \id_3$ is detected with a Schmidt number equal to 3 with a confidence interval of 2 standard deviations for a noise fraction $p=1/4$, while it is detected with a Schmidt number equal to 2 for a noise fraction $p=7/10$.

See also \cref{app:finitestat} for more details.

\section{CONCLUSIONS AND OUTLOOK}

In conclusion, we studied corollaries of the covariance matrix criterion for the Schmidt number~\cite{liu2022bounding}, which in some sense resemble well-known entanglement criteria~\cite{de2006separability,chen03,Rudolph2003} and can be seen as a generalization of them to a Schmidt-number witness.
Exploiting one of these corollaries, we have generalized the entanglement detection method of Ref.~\cite{imai2021bound} to witness the Schmidt number.
This method is based on the measurement of correlators of the form $\av{(U_a \otimes U_b)_i M \otimes M (U_a \otimes U_b)^\dagger_i}_\varrho$ for a given $M=\diag(\alpha_+,\dots,\alpha_+,\beta,\alpha_-,\dots,\alpha_-)$ and for random local unitaries $(U_a \otimes U_b)_i$ sampled from the Haar measure.
We have shown how such a method works in practice for bipartite states of equal dimensions $d=3,4$, showing also that it detects strictly more states than the other known criteria for the Schmidt number. 
We have also observed how our method detects paradigmatic example states after a finite sample of random correlation measurements.
Our method can be straightforwardly implemented in higher dimension, as long as the dimensions of the two systems are equal to each other. 
At the same time, the inequality that we have derived, \cref{eq:CoredCrit}, also works when the two local dimensions are different, which would enable the potential extension of the method based on randomized correlators to the multipartite case, where the notion of Schmidt number can be extended to that of Schmidt-number vector~\cite{Huber_2013}.
As we have observed, our criterion would detect strictly more states than typical Schmidt-number witnesses based on fidelities with respect to maximally entangled states, as well as criteria based on the $2$-norm of the correlation matrix, \cref{eq:HuberCrit}.
Its implementation with randomized measurements, however, relies on the results of Ref.~\cite{imai2021bound}, which provides concrete matrices $M$ such that the moments $\Corand^{(2)}$ and $\Corand^{(4)}$ in \cref{eq:Emoments} can be evaluated from a Bloch-sphere integral \eqref{eq:BLSPint}.
The question then arises how to generalize these results when the dimensions of the two parties are different, which would then allow for the generalization of our results for the detection of the full Schmidt-number vector in multipartite states.
This question is left for future investigation and provides an exciting research direction for implementing our method in the study of entanglement dimensionality in many-body states.

{\bf Acknowledgments.---}We thank Costantino Budroni, Julio de Vicente, and Nathaniel Johnston for discussions. This work is supported by the National Natural Science Foundation of China (Grants No. 12125402 and No. 11975026). Q.H. acknowledges the Beijing Natural Science Foundation (Z190005) and Innovation Program for Quantum Science and Technology (2021ZD0301702). M.H. acknowledges funding from the European Research Council (Consolidator grant 'Cocoquest' 101043705) and the EC grant  HyperSpace (101070168). O.G. acknowledges support by the Deutsche Forschungsgemeinschaft (DFG, German Research Foundation, project numbers 447948357 and 440958198), the Sino-German Center for Research Promotion (Project M-0294), the ERC (Consolidator Grant 683107/TempoQ) and the German Ministry of Education and Research (Project QuKuK, BMBF Grant No. 16KIS1618K). G.V. acknowledges financial support from the Austrian Science Fund (FWF) through the grants P 35810-N and P 36633-N (Stand-Alone).

{\it Note added:} While finishing this paper, we learned that related results were obtained in Ref. \cite{nikolai}.

\bibliography{references.bib}

\appendix

\onecolumngrid

\section{Moments of randomized correlators and Bloch-sphere integrals}\label{app:selfcontained}

In this section we summarize the derivation of the expressions in \cref{eq:ConstS} for the correlators of $su(d)$ generators averaged over the Bloch sphere and their relation with the second and fourth moments of the randomized correlators $\av{U_a M_a U_a^\dagger \otimes U_b M_b U_b^\dagger}_\varrho$. We essentially follow and summarize the results of Ref.~\cite{imai2021bound} and present it also here for making our work more self-contained. 
Let us expand the averaged correlators as
\be
\mathcal{S}_\varrho^{(m)} = N \int {\rm d} \hat{\vec \alpha} \int {\rm d} \hat{\vec \beta} \left[\av{\sigma_{\hat{\vec \alpha}}\otimes \sigma_{\hat{\vec \beta}}}_\varrho \right]^m = N \int {\rm d} \hat{\vec \alpha} \int {\rm d} \hat{\vec \beta} \left(\sum_{kl} \alpha_k \beta_l (\Cored)_{kl}\right)^m  .
\ee
Expanding the $m$-th power in terms of multinomial coefficients we can express the integrand as
\be
\left(\sum_{kl} \alpha_k \beta_l (\Cored)_{kl}\right)^m = \sum_{\sum_{kl} n_{kl}=m} \binom{m}{n_{11},n_{12},\dots,{n_{d^2-1 d^2-1}}} \prod_{kl=1}^{d^2-1} (\alpha_k \beta_l (\Cored)_{kl})^{n_{kl}} ,
\ee
and consequently we obtain
\be
\mathcal{S}_\varrho^{(m)} = N \sum_{\sum_{kl} n_{kl}=m} \binom{m}{n_{11},n_{12},\dots,{n_{d^2-1 d^2-1}}} \prod_{kl=1}^{d^2-1} (\Cored)_{kl}^{n_{kl}}  \int {\rm d} \hat{\vec \alpha} \prod_{k=1}^{d^2-1} \alpha_k^{a_k} \int {\rm d} \hat{\vec \beta} \prod_{l=1}^{d^2-1} \beta_l^{b_l} , 
\ee
where $a_k=\sum_l n_{kl}$ and $b_l=\sum_k n_{kl}$. Now, for fixed values of $a_k$ and $b_l$ the integral on the unit sphere is easily calculated from well-known expression in terms of $\Gamma$ functions. The only thing that remains is to check all possible combinations of $n_{kl}$ such that $\sum_{kl} n_{kl}=m$ and evaluate the integral for each such combination. 
In the case of $m=2$ there are only two possibilities, namely one value $n_{kl}=2$ and all the rest are zero or two values are equal to $1$, namely $n_{k_1 l_1}=n_{k_2 l_2}=1$ and the rest are zero. The second possibility leads to vanishing integrals, so only the first should be considered. In that case one gets 
\be
\mathcal{S}_\varrho^{(2)} = N \sum_{kl=1}^{d^2-1} (\Cored)_{kl}^{2} \frac{\sqrt{\pi^{d^2-1}}}{2{\mathrm \Gamma}(\frac{d^2+1}{2})} ,
\ee
where ${\mathrm \Gamma}(\cdot)$ is the $\Gamma$ function and the last factor comes from the integrations over the unit sphere.
Adjusting the normalization such that $\mathcal{S}_\varrho^{(2)}=1$ for pure product states we get $\mathcal{S}_\varrho^{(2)} = \tfrac{d^2}{(d-1)^2} \sum_{kl=1}^{d^2-1} (\Cored)_{kl}^{2}$ which is precisely the expression in \cref{eq:ConstS2}.

In the case of $m=4$ there are several more possibilities and the calculation is quite lengthy. Therefore, we refer the interested reader to Ref.~\cite{imai2021bound} for the detailed calculation that leads to 
the expression in \cref{eq:ConstS4}. Note that here we use a different normalization for the correlation matrix $\Cored$ with respect to Ref.~\cite{imai2021bound}, which leads to an additional factor of $d^m$ in the expressions of the $\mathcal{S}_\varrho^{(m)}$.

Let us now briefly summarize the calculations of $\Corand^{(2)}(M\otimes M)$ and $\Corand^{(4)}(M\otimes M)$ and the constraints on $M$ such that those are proportional to $\mathcal{S}_\varrho^{(2)}$ and $\mathcal{S}_\varrho^{(4)}$, respectively.
First of all, it has been shown in Ref.~\cite{Tran_2016} that the value of $\Corand^{(2)}(M\otimes M)$ is independent on the concrete eigenvalues (and eigenvectors) of $M$, as soon as $\tr(M)=0$. In particular, choosing the correct normalization $\tr(M^2)=d$  one obtains $\Corand^{(2)}(M\otimes M) = \tfrac{d^2}{(d^2-1)^2} \sum_i \epsilon_i^2 = \tfrac{1}{(d+1)^2}\mathcal{S}_\varrho^{(2)}$.
Concerning the fourth-order moment, one can expand the power inside the integral with the multinomial theorem. Imposing that $M$ is traceless one obtains similarly as for $\mathcal{S}_\varrho^{(m)}$:
\be
\Corand^{(m)}(M\otimes M) = \sum_{\sum_{kl} n_{kl}=m} \binom{m}{n_{11},n_{12},\dots,{n_{d^2-1 d^2-1}}} \prod_{kl=1}^{d^2-1} (\Cored)_{kl}^{n_{kl}}  \int {\rm d} U_a \prod_{k=1}^{d^2-1} \tr(U_a M U_a^\dagger \sigma_k)^{a_k} \int {\rm d} U_b \prod_{l=1}^{d^2-1} \tr(U_b M U_b^\dagger \sigma_l)^{b_l} , 
\ee
where $\sigma_k$ are elements of a $su(d)$ basis.
Again, for the case $m=4$ one has to check and sum up all possible combinations of $n_{kl}$ such that $\sum_{kl} n_{kl} = 4$.
As a result, one obtains that for a matrix of the form $M=\diag(\alpha_+,\dots,\alpha_+,\beta,\alpha_-,\dots,\alpha_-)$
and for properly chosen eigenvalues $\alpha_\pm$ and $\beta$ the fourth-order moment is given by
\be
\Corand^{(4)}(M\otimes M) = \frac{9(d-1)^2}{(d+1)^2\left(d^2+1\right)^2} \mathcal{S}_\varrho^{(4)} .
\ee
One concrete choice of eigenvalues is given, for the case of odd dimension $d$, by 
\be
\begin{aligned}
    \alpha_\pm &= \frac{\pm d - 2y +1}{\sqrt{(d-1)[(2y-1)^2 +d]}} , \\
    \beta_y &= - \sqrt{\frac{(d-1)(2y-1)^2}{(2y-1)^2 +d}} , \\
    \text{with} \quad y &= \frac 1 2 \left( 1 - \sqrt{1+ \frac{d+3+\sqrt{d^3+3d^2+d+3}}{d-2}} \right) .
\end{aligned}
\ee
A similar set of eigenvalues can be provided for the case of even $d$. See Ref.~\cite{imai2021bound} for the details.

\section{Boundary in $(\mathcal S^{(2)},\mathcal S^{(4)})$ for all states}\label{app:boundarystates}

\begin{figure}[h]
\centering
\includegraphics[width=0.48\textwidth]{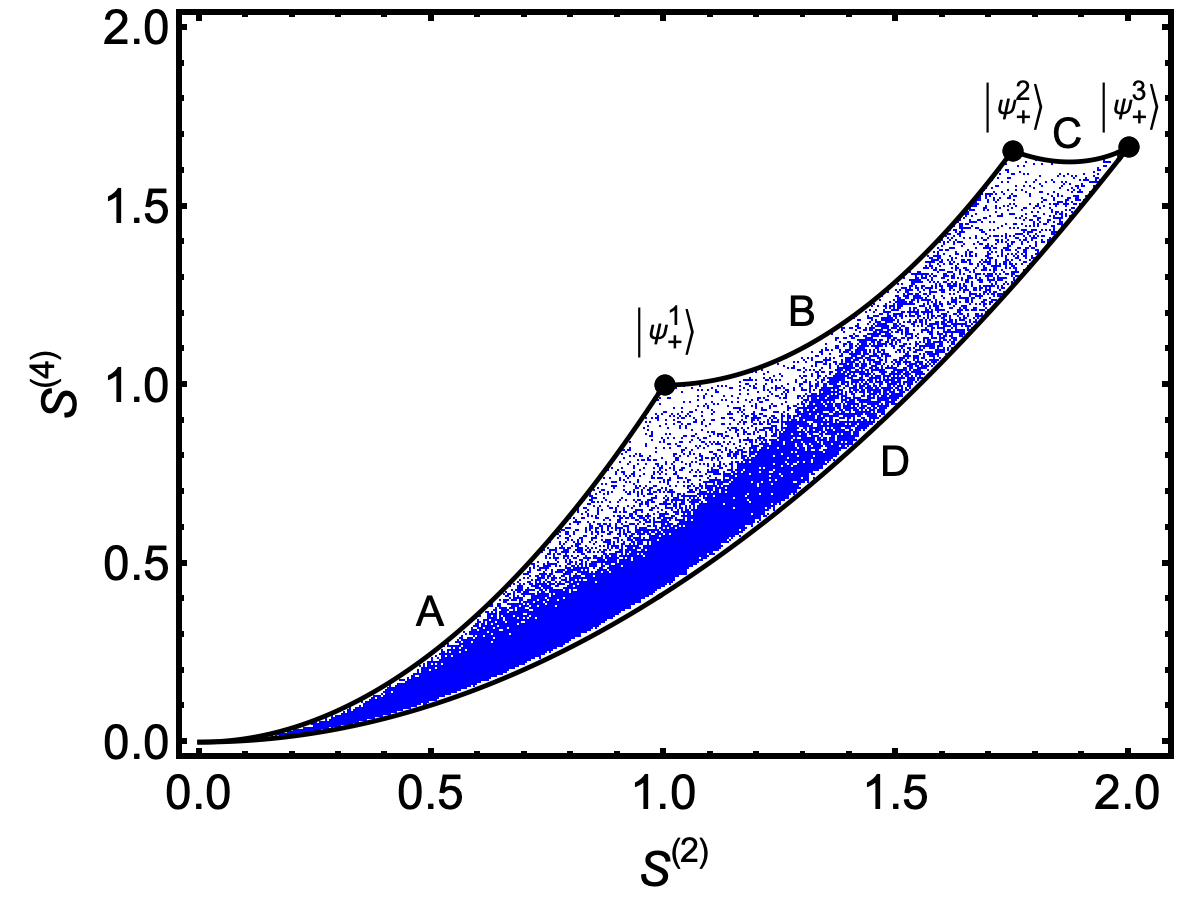}
\caption{External boundaries of the region in the $(\mathcal S^{(2)},\mathcal S^{(4)})$ plane reached by physical states. Internal points correspond to randomly sampled states, while the vertices correspond, respectively, to the maximally mixed state at the origin and pure states of increasing dimension $r=1,2,3$ at the other vertices. The boundary lines $A,B,C,D$ are reached by combination of states lying at the vertices (see text).}
\label{fig:d3randomsamples}
\end{figure}

Here we observe that the external boundaries 
depicted in \cref{fig:CriterionBoundariesD3} are reached by physical states and all the region within them is completely filled by physical states as well.  
A similar observation can be made for the $d=4$ figure, which we however omit here for simplicity.
In \cref{fig:d3randomsamples} we have plotted the boundary regions in the plane $(\mathcal S^{(2)},\mathcal S^{(4)})$, which are valid for all states, together with points corresponding to randomly sampled physical states. It can be observed how such random samples fill the whole region.

We can also observe what states lie on the boundaries. Curve $A$ corresponds to the family of mixed states $\rho_A(p)=p\ket{00}\bra{00}+\tfrac{1-p}{9}\mathbb{1}_{9}$ and is given by the function $y=x^2$ with $0\leq x \leq 1$.
Curve $B$ corresponds to the family of pure states $\ket{\psi_B(\lambda)}=\sqrt{\lambda}\ket{00}+\sqrt{1-\lambda}\ket{11}$ with $\tfrac 1 2 \leq \lambda \leq 1$ and is given by the function $y=\tfrac{7x^2}{6}-\tfrac{7x}{3}+\tfrac{13}{6}$ for $1 \leq x \leq \tfrac{7}{4}$.
Curve $C$ corresponds to the family of pure states $\ket{\psi_C(\lambda)}=\sqrt{\lambda}\ket{00}+\sqrt{\lambda}\ket{11}+\sqrt{1-2\lambda}\ket{22}$ with $\tfrac 1 3 \leq \lambda \leq \tfrac 1 2$ and is given by the function $y=\tfrac{7x^2}{6}+(2-x)^{\tfrac{3}{2}}-\tfrac{23x}{6}+\tfrac{14}{3}$ for $\tfrac{7}{4} \leq x \leq 2$.
Curve $D$ corresponds to the family of mixed states $\rho_D(p)=p\kb{\Psi_+^3}{\Psi_+^3}+\frac{(1-p)}{9}\mathbb{1}_{9}$ and is given by
the function $y=\tfrac{5x^2}{12}$ for $0 \leq x \leq 2$.
In the families of mixed states above, the parameter $p$ runs from $0$ to $1$.

\section{Analytical expression of Schmidt-number curves}\label{app:KKTsolutions}

Here, we are interested in finding the analytical solution to the optimization problem 
\begin{equation}
\begin{aligned}\label{eq:minProbApp}
\min_{\left\{\epsilon_i\right\}} & \sum_i \epsilon_i^4 \\
\text { subject to } & \epsilon_i \geq 0 \\
& \sum_i \epsilon_i^2=\frac{(d-1)^2}{d^2} \mathcal{S}^{(2)} := X \\
& \sum_i \epsilon_i \leq r-\frac{1}{d} ,
\end{aligned}
\end{equation}
which arises from \cref{eq:SchmidtProb} in the case of minimization.
One can observe that the resulting curves are the nontrivial ones for detecting the Schmidt number (see \cref{fig:CriterionBoundariesD3,fig:CriterionBoundariesD4} and \cref{app:boundarystates}).
The Lagrangian function for the problem \eqref{eq:minProbApp} is
\begin{equation}
\mathcal L(\{\epsilon_i\}, \{\lambda_i\}, \nu)=\sum_{i=1}^{d^2-1} \epsilon_i^4-\sum_{i=1}^{d^2-1} \lambda_i \epsilon_i+\lambda_0 \left(\sum_i \epsilon_i-r+\frac{1}{d}\right)+\nu\left(\sum_i \epsilon_i^2-X\right)
\end{equation}
where $\lambda_i\geq 0$ and $\nu$ are Lagrange multipliers. The Karush-Kuhn-Tucker conditions associated to the problem are
\begin{equation}
\begin{cases}
\frac{\partial \mathcal L}{\partial \epsilon_i}=4 \epsilon_i^3-\lambda_i+\lambda_0+2 \nu \epsilon_i=0 \quad \text{for} \quad 1\leq i\leq d^2-1 & \text { (stationarity) } \\ 
\sum_{i=1}^{d^2-1}\lambda_i \epsilon_i=0 & \text { (complementary slackness) } \\ 
\lambda_0\left(\sum_i \epsilon_i-r+\frac{1}{d}\right)=0 & \\ 
\sum_i \epsilon_i \leq r-\frac{1}{d} & \text { (primal feasibility) } \\ 
\sum_i \epsilon_i^2=X & \\ 
\lambda_i \geq 0 \quad \text{for} \quad 0\leq i\leq d^2-1 . & \text { (dual feasibility) }
\end{cases}.
\end{equation}
Let us now calculate the optimal points coming from these conditions. These are extremal points for the problem, not necessarily global minima. However, we see that these conditions are satisfied only by a finite number of points. Therefore, we check all of them and observe which ones provide the global minimum.
Let us distinguish two cases: (i) $\lambda_0>0$ and (ii) $\lambda_0=0$.

(i) Due to complementary slackness, we have $\sum_i \epsilon_i = r-\tfrac{1}{d}$. At the same time, if $\lambda_i=0$ for some $i>0$, then for the corresponding index we have $\epsilon_i \geq 0$ and $4 \epsilon_i^3+\lambda_0+2 \nu \epsilon_i=0$ due to stationarity. This equation has at most two positive solutions, which we call $\epsilon_i=s$ and $\epsilon_i=l$ (for ``small'' and ``large'').
If instead $\lambda_i>0$, then $\epsilon_i=0$ is a solution, and stationarity implies $\lambda_i=\lambda_0$. 
Thus, in total for the case (i) we have three possible values of each $\epsilon_i$, namely $\epsilon_i=\{0,s,l\}$.

Let us now check within all the possible (finite) combinations of $\epsilon_i=\{0,s,l\}$, which one gives the global minimum.
Assume the total number of $\epsilon_i=s$ to be $n_s$ and the total number of $\epsilon_i=l$ to be $n_l$. The remaining $\epsilon_i$'s are all zero. We now prove that the minimum is achieved for $n_s=1$. For this, we consider the following minimization problem:
\begin{equation}
\begin{aligned}
\min_{n_s, n_l \in \mathbb N} &  \left[ s^4 n_s + l^4 n_l \right], \\
\text { subject to } & s n_s+ l n_l=A>0, \\
& s^2 n_s +l^2 n_l =B>0 .
\end{aligned}
\end{equation}
First, let us express $s$ and $l$ in terms of the other parameters:
\begin{equation}
\begin{aligned}\label{eq:slexpsApp}
s&=\frac{A}{n_s+n_l}-\sqrt{\frac{-A^2 n_l+n_s n_l B+n_l^2 B}{n_s(n_s+n_l)^2}}, \\
l&=\frac{A}{n_s+n_l}+\frac{n_s \sqrt{\frac{n_l\left(-A^2+(n_s+n_l) B\right)}{n_s(n_s+n_l)^2}}}{n_l} .
\end{aligned}
\end{equation}
From $s,l>0$ we have 
\begin{equation}
\begin{aligned}
&0<b<\frac{A^2}{B}, \\
&a \geq \frac{A^2-b B}{B}
\end{aligned} 
\end{equation}
Substituting the expressions in \cref{eq:slexpsApp} for $s,l$ we get 
\begin{equation}
n_s s^4+ n_l l^4=\frac{\frac{\left(\sqrt{n_s n_l \left(B (n_s+n_l)-A^2\right)}+A n_l\right)^4}{n_l^3}+n_s \left(A-\sqrt{\frac{n_l \left(B
   (n_s+n_l)-A^2\right)}{n_s}}\right)^4}{(n_s+n_l)^4}
\end{equation}
Assume $n_s+n_l=q$ is a fixed number. This means the total number of nonzero singular values is fixed, and we aim to prove the target function decreases when we substitute a small singular value with a large one. Treating $n_s$ as a positive real variable we get
\begin{equation}
n_s s^4+n_l l^4=\frac{n_s \left(A-\sqrt{\frac{(n_s-q) \left(A^2-B q\right)}{n_s}}\right)^4+\frac{\left(\sqrt{n_s (n_s-q) \left(A^2-B
   q\right)}-n_s A+A q\right)^4}{(q-n_s)^3}}{q^4},
\end{equation}
where $n_s<q$, $B q<n_s B+A^2$ and $A^2\leq B q$, which are a consequence of $s,l>0$. The derivative with respect to $n_s$ of the above function is always non-negative, namely $\frac{\partial (n_s s^4+n_l l^4)}{\partial n_s}\geq 0$. Here $s,l$ and $n_l$ are all dependent on $n_s$. Thus $n_s$ should take the smallest positive integer, that is $n_s=1$. In this case, correspondingly we get $n_l=\lceil\frac{A^2}{B}-1\rceil$ ($0<\mathrm{B}<\mathrm{A}^2$). Note that the case of $s=l$ is also considered, which is equivalent to the case that all nonzero singular values are equal.

Finally, we come back to our original optimization problem, which corresponds to
$A=r-\frac{1}{d}$ and $B=\frac{(d-1)^2}{d^2} \mathcal{S}^{(2)}$. The total number of singular values is $d^2-1$. Taking the different values of $n_l=\lceil\frac{A^2}{B}-1\rceil=1,\cdots,d^2-2$ we get
\begin{equation}
\frac{(d r-1)^2}{(d-1)^2(n_l+1)} \leq \mathcal{S}^{(2)}<\frac{(d r-1)^2}{(d-1)^2 n_l} ,
\end{equation}
which means that we have found the minimum for these ranges of values of $\mathcal{S}^{(2)}$.
Such a minimum is obtained by substituting $n_s,n_l,A,B$ in $\mathcal{S}^{(4)}$:
\begin{equation}
\mathcal{S}^{(4)}=\frac{2\left(\left(\sqrt{n_l(n_l+1) \mathcal{S}^{(2)}-n_l\left(\frac{d r-1}{d-1}\right)^2}-\frac{d r-1}{d-1}\right)^4+\frac{\left(\sqrt{n_l(n_l+1) \mathcal{S}^{(2)}-n_l\left(\frac{d r-1}{d-1}\right)^2}+n_l \frac{d r-1}{d-1}\right)^4}{n_l^3}\right)}{3(n_l+1)^4}+\frac{1}{3}\left(\mathcal{S}^{(2)}\right)^2 .
\end{equation}
Thus, in total we have found the minimum in the range
\begin{equation}
\frac{(d r-1)^2}{(d-1)^2\left(d^2-1\right)} \leq \mathcal{S}^{(2)}<\frac{(d r-1)^2}{(d-1)^2} .
\end{equation}
What is still missing is the range $0 \leq \mathcal{S}^{(2)}<\frac{(d r-1)^2}{(d-1)^2\left(d^2-1\right)}$, because the maximum value for $n_l$ is $d_1^2-2$.

For this range, the minimum is achieved when $\lambda_0=0$, that is what we referred to as case (ii).
In this case, when also $\lambda_i=0$ for some $i>0$, using complementary slackness we have $\epsilon_i \geq 0$. On the other hand, the case $\lambda_i>0$ is not allowed for any $i>0$ since it would imply 
$\epsilon_i=0$, which is not a valid solution for stationarity, namely $4 \epsilon_i^3-\lambda_i+\lambda_0+2 \nu \epsilon_i=0$. Therefore, we must have $\lambda_i=0$ for all $i$, and $\epsilon_i$ satisfy $4 \epsilon_i^3+2 \nu \epsilon_i=0$, which means $\epsilon_i=\sqrt{-\frac{\nu}{2}}$ or $\epsilon_i=0$. Assume the total number of nonzero $\epsilon_i$ is $n$ where $n=1,\cdots,d^2-1$. Let us then consider the minimization:
\begin{equation}
\begin{array}{cl}
\min & n \epsilon_i^4 \\
\text { subject to } & n \epsilon_i\leq A \\
& n \epsilon_i^2=B
\end{array}
\end{equation}
where $A=r-\frac{1}{d}$ and $B=\frac{(d-1)^2}{d^2} \mathcal{S}^{(2)}$. From $n \epsilon_i^2=B$ we know that $\epsilon_i=\sqrt{\frac{B}{n}}$. The constraint $n \epsilon_i \leq A$ becomes $n\leq \frac{A^2}{B}$ and the target function becomes $\frac{B^2}{n}$. The minimum of $\frac{B^2}{n}$ is achieved when $n$ is the largest possible integer, i.e., $n=d^2-1$. 
In this case we indeed get $0 \leq \mathcal{S}^{(2)} \leq \frac{(d r-1)^2}{(d-1)^2\left(d^2-1\right)}$ and the corresponding minimum is 
\begin{equation}
\mathcal{S}^{(4)}=\frac{d^2+1}{3 (d^2-1)}\left(\mathcal{S}^{(2)}\right)^2 .
\end{equation}

\section{Detection of maximally entangled state with finite statistics}\label{app:finitestat}

Here we test our method in a paradigmatic example with a finite statistical sample of local unitaries.
We consider $d=3$, and $\ket{\Psi_{+}^3}$ as the target measured state and simulate numerically the statistics that would ideally occur with a finite sample of randomized correlators. For our target state, the ideal second- and fourth-order integrals are $\mathcal{S}_{\Psi_{+}^3}^{(2)}=2$ and $\mathcal{S}_{\Psi_{+}^3}^{(4)}=\frac{5}{3}$. We take a random numerical sample of, respectively, $N_{\rm tot}=\{10^3,10^4,10^5\}$ local unitaries from the Haar measure and estimate the moments of the randomized correlators as
\be
\begin{aligned}
\hat{\mathcal{S}}_{\Psi_{+}^3}^{(2)} &= (d+1)^2 \tfrac{1}{N_{\rm tot}} \sum_{i} \bra{\Psi_{+}^3} (U_a \otimes U_b)_i M \otimes M (U_a \otimes U_b)^\dagger_i \ket{\Psi_{+}^3}^2 := (d+1)^2 \tfrac{1}{N_{\rm tot}} \sum_{i} x_i^2(U_a,U_b), \\ 
\hat{\mathcal{S}}_{\Psi_{+}^3}^{(4)} &= \tfrac{(d+1)^2\left(d^2+1\right)^2}{9(d-1)^2} \tfrac{1}{N_{\rm tot}} \sum_{i} x_i^4(U_a,U_b) ,
\end{aligned}
\ee
and for independent and Haar-distributed points we get
\begin{equation}
\begin{aligned}
\mu\left(\hat{\mathcal{S}}_\varrho^{(2)}\right) & =(d+1)^2 \frac{1}{N} \sum_{i=1}^N \mu\left(x_i^2\right)=(d+1)^2 \mu\left(x^2\right) = (d+1)^2 \int x^2(U_a,U_b) \mathrm{d} U_a \mathrm{d} U_b = \mathcal{S}_\varrho^{(2)} \\
\mu\left(\hat{\mathcal{S}}_\varrho^{(4)}\right) & =\frac{(d+1)^2\left(d^2+1\right)^2}{9(d-1)^2} \frac{1}{N} \sum_{i=1}^N \mu\left(x_i^4\right)=\frac{(d+1)^2\left(d^2+1\right)^2}{9(d-1)^2} \mu\left(x^4\right) =  \mathcal{S}_\varrho^{(4)} ,
\end{aligned}
\end{equation}
i.e., the estimators are unbiased. 
Then, we can estimate the error by means of the standard deviation of the estimators, which are given by
\be
\begin{aligned}
\sigma^2\left(\hat{\mathcal{S}}_\varrho^{(2)}\right) &= (d+1)^4 \frac{1}{N^2} \sum_{i=1}^N \sigma^2\left(x_i^2\right)=\frac{(d+1)^4}{N} \sigma^2\left(x^2\right)=\frac{(d+1)^4}{N_{\rm tot}} \left( \Corand^{(4)} - (\Corand^{(2)})^2 \right) , \\
\sigma^2\left(\hat{\mathcal{S}}_\varrho^{(4)}\right) & =\frac{(d+1)^4\left(d^2+1\right)^4}{81(d-1)^4} \frac{1}{N^2} \sum_{i=1}^N \sigma^2\left(x_i^4\right)=\frac{(d+1)^4\left(d^2+1\right)^4}{81(d-1)^4 N} \sigma^2\left(x^4\right)=\frac{(d+1)^4\left(d^2+1\right)^4}{81(d-1)^4 N_{\rm tot}} \left(\Corand^{(8)} - (\Corand^{(4)})^2 \right) ,
\end{aligned}
\ee
Knowing the analytical values of $\Corand^{(2)}$ and $\Corand^{(4)}$ we can get
\begin{equation}
\begin{aligned}
\mu\left(x^2\right)&=\frac{1}{8}=0.125,\\
\mu\left(x^4\right)&=\frac{3}{80}=0.0375,\\
\sigma^2\left(x^2\right)=\mu\left(x^4\right)-\mu^2\left(x^2\right)&=\frac{7}{320}=0.021875 .
\end{aligned}
\end{equation}
However, we do not have the analytical expression for $\Corand^{(8)} = \mu(x^8)$ and thus we calculate it numerically with $2\times 10^6$ random samples of $x$. For that we get the value $\mu(x^8) \approx 0.007925$ and correspondingly $\sigma^2(x^4) \approx 0.006520$.

\begin{figure}[h]
\centering
\includegraphics[width=0.48\linewidth]{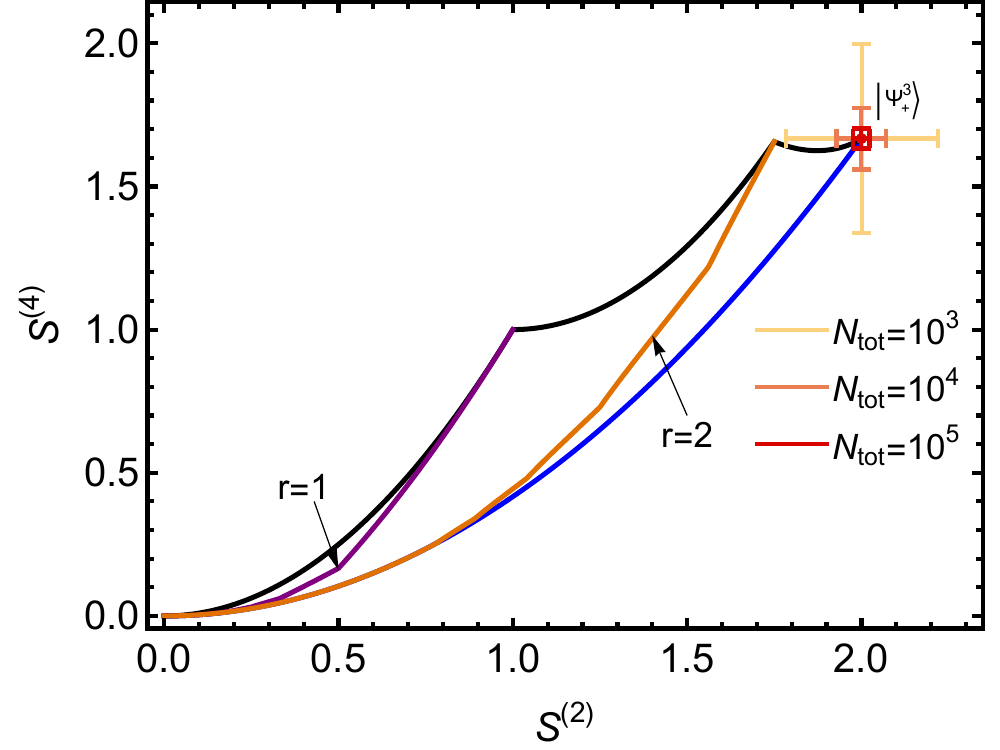}
\caption{Plot of the estimation of $\hat{\mathcal{S}}_{\Psi_{+}^3}^{(2)}$ for the $d=3$ maximally entangled state with a $3\sigma$ uncertainty region together with the boundaries for the different Schmidt number $r=1,2,3$. The largest error bar is for a sample of $N_{\rm tot}=10^3$, the middle is for $N_{\rm tot}=10^4$ and the smallest is for $N_{\rm tot}=10^5$.}\label{fig:sampleAppD1}
\end{figure}

As a result, mean values with an uncertainty given by 3 standard deviations (corresponding to $99.73\%$ confidence level) are
\begin{equation}
\begin{array}{lll}
\hat{\mathcal{S}}_{\Psi_{+}^3}^{(2)}=2 \pm 0.224, & \hat{\mathcal{S}}_{\Psi_{+}^3}^{(4)}=1.667 \pm 0.340, & N_{\rm tot}=10^3, \\
\hat{\mathcal{S}}_{\Psi_{+}^3}^{(2)}=2 \pm 0.071, & \hat{\mathcal{S}}_{\Psi_{+}^3}^{(4)}=1.667 \pm 0.108, & N_{\rm tot}=10^4, \\
\hat{\mathcal{S}}_{\Psi_{+}^3}^{(2)}=2 \pm 0.022, & \hat{\mathcal{S}}_{\Psi_{+}^3}^{(4)}=1.667 \pm 0.034, & N_{\rm tot}=10^5 .
\end{array}
\end{equation}
As shown in the plot in \cref{fig:sampleAppD1}, Schmidt number $3$ is certified in all cases. Therefore, a sample size of $10^3$ is already enough.

\begin{figure}[h]
\centering
\includegraphics[width=0.48\textwidth]{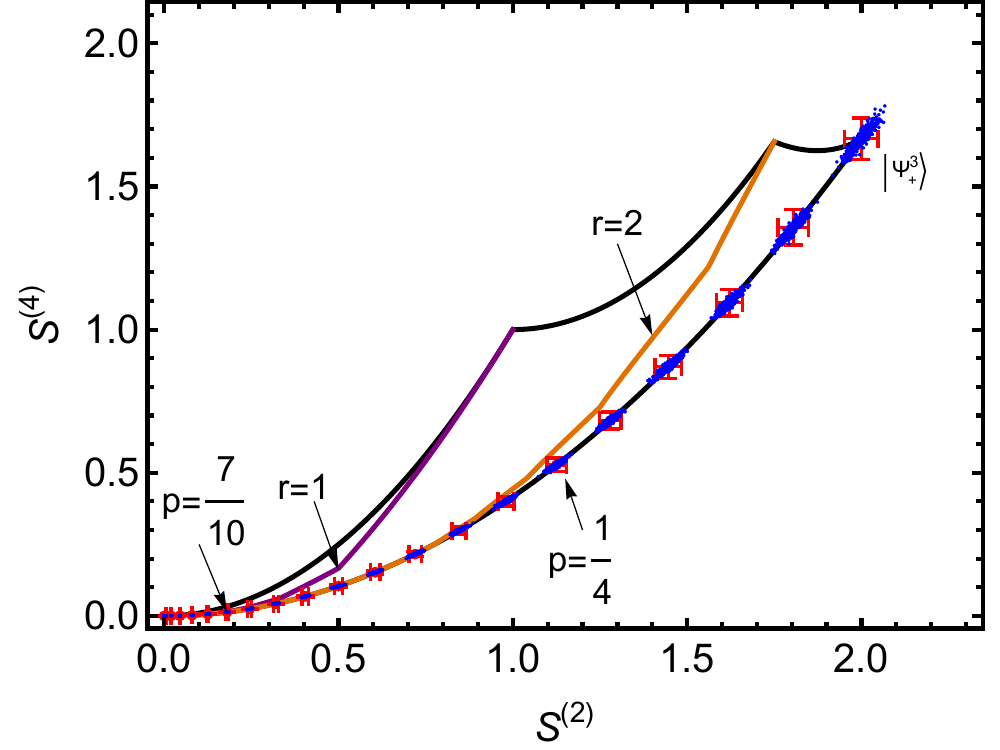}
\caption{Sampling points (blue) and confidence intervals of two standard deviations (red) with respect to different noise fractions in $\varrho_p$. The noise fractions $p=7/10$ and $p=1/4$ marked on the error bars represent the thresholds of $p$ for Schmidt numbers $2$ and $3$, respectively.}
\label{fig:ErrorBarProbD3}
\end{figure}

Similarly, one can calculate the white-noise tolerance of the method, for example, with a sample of $N_{\rm tot}=10^4$ random unitaries. In this case we get that the state 
$\varrho_p := (1-p)\ketbra{\Psi_{+}^3}+p\frac{\id_3 \otimes \id_3} 9$ is detected with a Schmidt number equal to 3 with a confidence interval of 2 standard deviations for a noise fraction $p=1/4$, while it is detected with a Schmidt number equal to 2 for a noise fraction $p=7/10$, as shown in \cref{fig:ErrorBarProbD3}.

\end{document}